\documentclass{aastex6}
\newcommand{\droh}{DR21(OH)}
\newcommand{\Blos}{\ensuremath{B_{\text{los}}}}
\newcommand{\meth}{CH$_3$OH}
\newcommand{\jybeam}{Jy~beam$^{-1}$}
\newcommand{\kmS}{km~s$^{-1}$}

\shorttitle{44 GHz Methanol maser Zeeman Effect toward DR21(OH)}
\shortauthors{Momjian \& Sarma}

\begin{document}

\title{The Zeeman Effect in the 44 GHz \\ Class I Methanol Maser Line toward DR21(OH)}

\author{E.~Momjian\altaffilmark{1} and A.~P.~Sarma\altaffilmark{2}}

\altaffiltext{1}{National Radio Astronomy Observatory, P.O~Box 0, Socorro, NM 87801; emomjian@nrao.edu}
\altaffiltext{2}{Physics Department, DePaul University, 2219 N.~Kenmore Ave., Byrne Hall 211, Chicago, IL 60614; asarma@depaul.edu}

\begin{abstract}
We report the detection of the Zeeman effect in the 44 GHz Class I methanol maser line toward the star forming region DR21(OH). In a 219~Jy~beam$^{-1}$ maser centered at an LSR velocity of 0.83~km s$^{-1}$, we find a 20-$\sigma$ detection of $zB_{\text{los}} = 53.5 \pm 2.7$~Hz. If 44 GHz methanol masers are excited at $n \sim 10^{7-8}$~cm$^{-3}$, then the $B~vs.~n^{1/2}$ relation would imply from comparison with Zeeman effect detections in the CN($1-0$) line toward DR21(OH) that magnetic fields traced by 44 GHz methanol masers in \droh\ should be $\sim$10~mG. Together with our detected $zB_{\text{los}} = 53.5$~Hz, this would imply that the value of the 44 GHz methanol Zeeman splitting factor $z$ is $\sim$5~Hz~mG$^{-1}$. Such small values of $z$ would not be a surprise, as the methanol molecule is non paramagnetic, like H$_2$O. Empirical attempts to determine $z$, as demonstrated, are important because currently there are no laboratory measurements or theoretically calculated values of $z$ for the 44 GHz \meth\ transition. Data from observations of a larger number of sources are needed to make such empirical determinations robust.

\end{abstract}

%% Keywords

\keywords{masers --- polarization --- ISM: individual objects (DR21OH) --- ISM: magnetic fields --- ISM: molecules --- stars: formation}

\section{Introduction} \label{intro}

Class I methanol (\meth) masers are excellent signposts of high mass star forming regions. They are generally found in outflows where the collisional shocks pump the maser transitions \citep*[see, e.g.,][]{leurini+2016}. The Zeeman effect in Class I \meth\ masers offers unprecedented opportunities for measuring magnetic field strengths at high angular resolution in such star forming regions. It has long been acknowledged that magnetic fields must play an important role at various stages in the star forming process, but the nature of their role remains a matter of debate \citep[see, e.g.,][and references therein]{masson+2016}. Observations of magnetic fields in a range of environments are critical to provide useful inputs to theory \citep[see, e.g.,][and references therein]{crutcher+2012}. 

DR21(OH) is a massive star formation region located in the Cygnus-X molecular cloud complex \citep[see, e.g.,][]{odenwald+93, motte+2007}. 
Recent astrometric maser measurements put it at a distance of 1.50 kpc \citep{rygl+2012}. It is embedded in a 4~pc long filament that extends along the north-south direction. Herschel observations by \citet{henn+2012} show several dense high mass star forming cores in this filament, including at the location of \droh. On large scales, single dish polarimetric observations with the James Clerk Maxwell Telescope (JCMT) at $850~\micron$ and 20\arcsec\ angular resolution reveal uniform plane-of-sky magnetic fields that are oriented along the east-west direction, perpendicular to the filament \citep{brenda+2009}. However, high angular resolution ($\sim 1\arcsec$) polarimetric observations of the \droh\ core taken with the Smithsonian Millimeter Array (SMA) at $850~\micron$ reveal a complex magnetic field morphology in the plane of the sky, suggestive of a toroidal configuration \citep{girart+2013}. Indicators of ongoing star formation in \droh\ include strong outflows in CO and other molecular lines, and maser lines of several molecules \citep[see, e.g.,][and references therein]{zapata+2012}. 

\citet{kogan+98} observed 44 GHz Class I \meth\ masers in \droh\ and found them to be distributed in two elliptically-shaped regions. \citet{kurtz+2004}\ detected 17 such \meth\ maser spots at 44 GHz in two arc-shaped structures arranged symmetrically about a central source, suggestive of a bipolar outflow along the east-west direction. \citet{araya+2009}\ detected a total of 49 Class I \meth\ maser spots at 44 GHz in \droh, and found that the masers in both the eastern and western lobes are distributed in not one, but two arc-like structures. The redshifted masers are located predominantly in the western lobe, whereas blueshifted masers are predominantly in the eastern lobe \citep{araya+2009}.

In this paper, we report the detection of the Zeeman effect in the 44 GHz Class I \meth\ maser line toward \droh. In \S\ \ref{obsred}, we present details of the observations and data reduction. In \S\ \ref{analysis} we describe the analysis of the data, including the use of the new tool ZEMAN in AIPS for Zeeman effect data analysis. The results of our observations are presented and discussed in \S\ \ref{resdisc}. In \S\ \ref{conc}, we present our conclusions. 

\section{Observations and Data Reduction} \label{obsred}

The observations of the $7_{0}-6_1\, A^+$ Class I \meth\ maser emission line at 44 GHz toward \droh\ were carried out with the Karl G. Jansky Very Large Array
(VLA)\footnote{The National Radio Astronomy Observatory (NRAO) is a facility of the National Science Foundation operated under cooperative agreement by Associated Universities, Inc.} on 2012 April 25 in two consecutive 2\,hr sessions. The array was in C-configuration with a maximum baseline of 3.4\,km. The Wideband Interferometric Digital ARchitecture (WIDAR) correlator was configured to deliver a single 1\,MHz sub-band with dual polarization products (RR, LL) and 256 spectral channels. The resulting channel spacing was 3.90625\,kHz, which corresponds to 0.0266\,\kmS\ at the observed frequency. In addition to the target source, the calibrator J1331$+$3030 (3C286) was observed to derive the antenna-based amplitude gain factors in order to calibrate the absolute flux density scale. The uncertainty in the flux density calibration at the observed frequency, accounting for various observational parameters (e.g., weather, reference pointing, and elevation effects), is expected to be up to 10\%. Table\ \ref{tOP}\ summarizes the parameters of our VLA observations.

Data reduction, including calibration, deconvolution and imaging, was carried out using the Astronomical Image Processing System (AIPS; \citealt{greisen+2003}). After Doppler correcting the spectral-line data of \droh, the frequency channel with the brightest maser emission signal was split off, and self-calibrated first in phase, then in both phase and amplitude, and imaged in a succession of iterative cycles. The final self-calibration solutions were applied to the full spectral-line data set of \droh. Stokes $I$ and $V$ image cubes were constructed with a synthesized beamwidth of $0\, \rlap{\arcsec}.\, 63 \times 0\, \rlap{\arcsec}.\, 50$ at full width half maximum (FWHM) and at a position angle of 75.3$^{\circ}$.

\section{Analysis} \label{analysis}
 
The magnetic field strength in Zeeman effect observations is usually determined by fitting a numerical frequency derivative of the Stokes $I$ spectrum to the Stokes $V$ spectrum. The method has been described in detail most recently in \citet{sarma+2013}, but here we repeat the key points. For historical reasons, AIPS calculates the Stokes parameters as $I = $~(RCP + LCP)/2, and $V = $~(RCP $-$ LCP)/2, where RCP is right-circular polarization and LCP is left-circular polarization; we use the standard radio definition of RCP as a clockwise rotation of the electric vector when viewed along the direction of wave propagation. All quoted values of $I$ and $V$ in this paper refer to the AIPS-generated values defined above; even though $I$ and $V$ are both different by a factor of 1/2 from the conventional definition of Stokes $I$ and $V$, this does not affect Zeeman effect results since only the relative magnitudes of $V$ and $I$ matter, as discussed below. When the Zeeman splitting is much smaller than the width of the line ($\Delta \nu_Z \ll \Delta \nu$), the Stokes $V$ profile is usually fit to the equation \citep{troland+82, sault+90}
\begin{equation}
V = aI + \frac{b}{2}\, \frac{dI}{d\nu}, \label{eq1}
\end{equation}
since the observed $V$ spectrum may contain a scaled replica of the $I$ spectrum as a result of small calibration errors in RCP versus LCP. For the detection reported in this paper, the fit parameter $a$ that represents the scaled replica of the $I$ profile in the $V$ profile is $\lesssim 10^{-3}$. The fit parameter $b=zB\cos\, \theta$, where $B$ is the magnetic field strength, $\theta$ is the angle of the magnetic field to the line of sight, and $z$ is the Zeeman splitting factor for \meth. For all cases where the Zeeman splitting is much smaller than the linewidth, we can only determine the line of sight magnetic field strength, \Blos = $B\cos\, \theta$. Moreover, the Zeeman splitting factor $z$ has never been measured for the 44 GHz \meth\ transition. Extrapolation of the Land\'e $g$-factor from lab measurements of several \meth\ transitions near 25 GHz made by \citet{jen+1951}\ could certainly be used to make an estimate of $z$ for the 44 GHz \meth\ maser transition. However, such an extrapolation usually gives a value for the magnetic field that is at least an order of magnitude larger than expected (see, e.g., \citealt{vlemmings+2011, momjian+2012}). Therefore, we will leave our results in terms of $z\Blos$ (in units of Hz).

The numerical frequency derivative of Stokes $I$ was fitted to Stokes $V$ using the newly available AIPS task ZEMAN \citep{greisen+2015}. The advantage of this task is that it allows multiple Gaussian components in $I$ to be fitted simultaneously to $V$ with different values of $b$. In other words, each Gaussian component can be fitted for a different \Blos\ value. The fitting of Gaussian components in $I$ was carried out using the AIPS task XGAUS, to which several interactive enhancements have been added recently \citep{greisen+2015}. In summary, AIPS now provides the full capability for an end-to-end processing of Zeeman observations. Also convenient is that publication quality figures such as those presented in this paper can be generated in AIPS using the new task XG2PL. 

\section{Results and Discussion} \label{resdisc}

\begin{figure}[htb!]
\epsscale{0.5}
\plotone{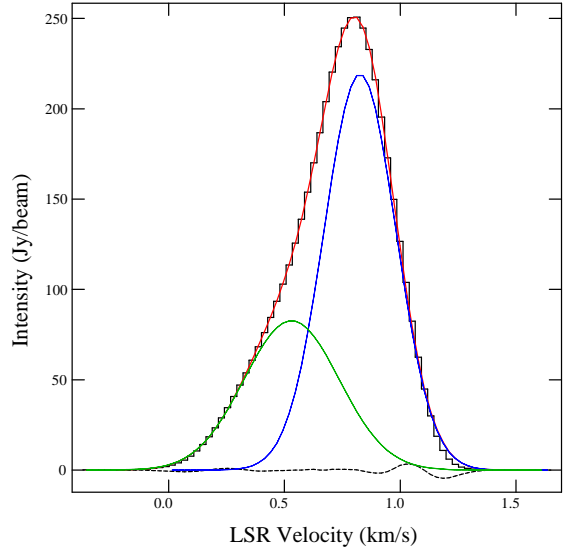}
\caption{Stokes $I$ profile (\textit{black histogram-like line}) toward the maser in \droh\ for which we report the Zeeman effect. The blue and green curves show the Gaussian components that we fitted to the Stokes $I$ profile (components 1 and 2 in Table~\ref{tGauss}, respectively), the red curve shows the sum of the two Gaussian component profiles, and the black dotted curve shows the residuals from the Gaussian fit. \label{fig1}}
\end{figure}

Figure~\ref{fig1} shows the Stokes $I$ profile toward the 44 GHz Class I \meth\ maser in \droh\ in which we detected significant Zeeman splitting. We fitted two Gaussian components to the Stokes $I$ profile. The intensity, center velocity, and full width at half maximum (FWHM) velocity linewidth of these two components are given in Table~\ref{tGauss}. Figure~\ref{fig1} also shows the two fitted Gaussian components (\textit{blue and green curves}; components 1 and 2 in Table~\ref{tGauss} respectively), together with their sum (\textit{red curve}) and the residuals (\textit{dotted black curve}) from the Gaussian fit. The fits essentially reveal two masers blended in velocity, with a 219~\jybeam\ maser centered at 0.83~\kmS, and a 82.7~\jybeam\ maser centered at 0.53~\kmS, with FWHM velocity linewidths $\sim$0.5~\kmS. The residuals suggest an excellent fit, although there is likely a third, very low-level component in Stokes $I$ between 1.0--1.3~\kmS. 

\begin{figure}[htb!]
\epsscale{0.5}
\plotone{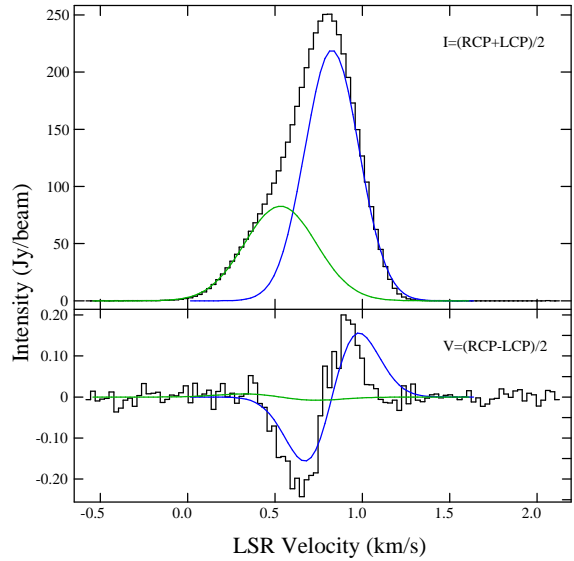}
\caption{Stokes $I$ (\textit{upper panel - black histogram-like line}) and Stokes $V$ (\textit{lower panel - black histogram-like line}) profiles toward the maser shown in Figure~\ref{fig1}. Again, the blue and green curves in the upper panel show the Gaussian components that we fitted to the Stokes $I$ profile (components 1 and 2 in Table~\ref{tGauss}, respectively). The blue curve in the lower panel is the derivative of the blue-colored Gaussian component in the upper panel, scaled by $z$\Blos$ = 53.5 \pm 2.7$~Hz, and the green curve in the lower panel is the derivative of the green-colored Gaussian component in the upper panel, scaled by $z$\Blos$ = -9.2 \pm 9.3$~Hz. \label{fig2}}
\end{figure}

\begin{figure}[htb!]
\epsscale{0.5}
\plotone{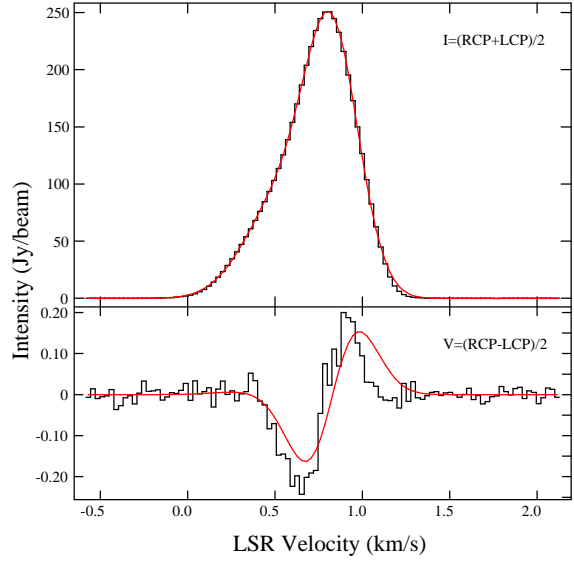}
\caption{Stokes $I$ (\textit{upper panel - black histogram-like line}) and Stokes $V$ (\textit{lower panel - black histogram-like line}) profiles toward the maser shown in Figure~\ref{fig1}. The red curve in the upper panel is the sum of the blue- and green-colored Gaussian components (shown in Figure \ref{fig1}\ and \ref{fig2}) that we fitted to the Stokes $I$ profile. The red curve superposed on the Stokes $V$ curve in the lower panel is the sum of the blue- and green-colored curves shown in the lower panel of Figure~\ref{fig2}; that is, it is the sum of the scaled derivatives of the Gaussian components fitted to the Stokes $I$ profile, where each of the two derivatives has been appropriately scaled by the value of $z\Blos$ given in Table~\ref{tGauss} and also in the caption to Figure~\ref{fig2}. \label{fig3}}
\end{figure}

Figure~\ref{fig2} shows the Stokes $I$ (\textit{upper panel -- black histogram-like line}) and $V$ (\textit{lower panel -- black histogram-like line}) profiles toward the \meth\ maser shown in Figure~\ref{fig1}. The blue and green curves in the upper panel are the two Gaussian components that we fitted to the Stokes $I$ profile. The blue curve in the lower panel is the derivative of the blue-colored Gaussian component in the upper panel (component 1 in Table~\ref{tGauss}), scaled by $z$\Blos$ = 53.5 \pm 2.7$~Hz, and the green curve in the lower panel is the derivative of the green-colored Gaussian component in the upper panel (component 2 in Table~\ref{tGauss}), scaled by $z$\Blos$ = -9.2 \pm 9.3$~Hz. This means that the blue-colored component shows a significant Zeeman detection at the level of 20-$\sigma$, whereas there is no significant detection in the green-colored component. By convention, a positive value for \Blos\ means that the line of sight magnetic field is pointing away from the observer. We have chosen to leave our results in $z\Blos$, because the value of $z$ for the 44 GHz transition of \meth\ has not been measured so far (but see \S\ \ref{analysis}). For clarity, we have shown the sum of the two components in a separate figure; Figure~\ref{fig3} also shows the Stokes $I$ and $V$ profiles as shown in Figure~\ref{fig2}, but this time with the sum of the Gaussians overlaid on $I$ (\textit{upper panel -- red curve}) and the sum of the appropriately scaled derivatives of $I$ overlaid on $V$ (\textit{lower panel -- red curve}). It is clear in Figure~\ref{fig3} that the right wing of the Stokes $V$ profile (1.0--1.3 \kmS) is systematically displaced from the profile of the scaled derivative. One reason for this discrepancy could be the third low-level component in Stokes $I$ mentioned above that is likely present between 1.0--1.3 \kmS. It is possible that this low-level feature is systematically affecting the Stokes $V$ profile. Another possibility is that the Stokes $V$ profile has undergone narrowing so that it is no longer strictly proportional to the derivative of the Stokes $I$ profile. Such narrowing was also observed and discussed by \citet{vlemmings+2001}. In such cases, derived $B$-fields could be off by a factor of 2, as estimated by \citet{vlemmings+2001}.

Our detection of the Zeeman effect in \droh\ is in the second strongest 44 GHz \meth\ maser spot within the field and the velocity span of our observations. Our observations were aimed at detecting the Zeeman effect in such masers following the initial discoveries in the 36 and 44 GHz Class I \meth\ masers \citep{sarma+2009, momjian+2012}. Therefore, we carried out the observations by maximizing the amount of time on the target source within the observing sessions to secure detections in Stokes $V$. As a result, we did not spend time on phase calibrator scans to get absolute positions for the masers. We were, however, able to find absolute positions for the masers by doing fringe-rate mapping on the strongest maser in the field (323 \jybeam) using the AIPS task FRMAP. The positions of the 24 maser spots ($\ge 0.3$~\jybeam) that we detected in \droh, together with their intensities, center velocities, and velocity linewidth at FWHM are listed in Table~\ref{tGauss_allmasers}. The distribution of the 24 maser spots in our observations listed in Table~\ref{tGauss_allmasers} indicates that we have recovered the double arc structure in the redshifted western lobe described in \citet{araya+2009}. Note that our velocity coverage would not have allowed us to detect any of the masers in the blueshifted eastern lobe. Of interest is that our strongest maser is in the inner arc of the western lobe, whereas the strongest maser in \citet{araya+2009} was in the outer arc. There is reasonable agreement in position between their strongest maser (labeled as maser 3 in their paper) and the second strongest maser in our observations (i.e., the maser in which we have the Zeeman detection), and likewise, their second strongest maser and our strongest maser also agree in position. However, there appears to be a mismatch of about 0.5~\kmS\ in center velocities between our observations and those of \citet{araya+2009}; note that their channel width, which is 0.66~\kmS, is significantly coarser compared to our observations. Such discrepancies preclude a direct comparison of intensities between their observations and ours, and raise the necessity for a coordinated interferometric study of the variability of Class I \meth\ masers, which to our knowledge has not been carried out so far. We did not detect the Zeeman effect in the strongest maser ($\sim$300~\jybeam) in our observations. This is not entirely surprising given how complex the plane-of-sky magnetic field is known to be in the DR21(OH) core from polarimetric observations \citep{girart+2013}. Our strongest maser is located about 11\arcsec\ from the second strongest maser in which we detected the Zeeman effect, and the SMA polarimetric observations of \citet{girart+2013} reveal plane-of-sky magnetic field vectors rotated by 90\arcdeg\ over 10\arcsec\ scales. Therefore, it is easy to imagine the field having a component along the line of sight at the position of our second strongest maser in the outer arc of the western lobe and changing direction so that it has no line of sight component at the position of our strongest maser 11\arcsec\ away. All the other masers in our observations are under 20~\jybeam, and lack the sensitivity to detect the Zeeman effect. 

In all Zeeman observations, it is worth considering whether the detected signal could be due to non-Zeeman effects. Instrumental effects can become an issue in extended sources where a velocity gradient across the source could mimic a Zeeman detection, especially in single dish observations. Such a possibility appears unlikely in maser observations with interferometers since masers are point sources that are confined to a narrow velocity range. If the detected signal were a false one due to instrumental polarization, especially when the source is off the peak of the primary beam, then such a signal would be dependent on the parallactic angle. Therefore, the most effective way to check for such false signatures of instrumental polarization is to check for consistent results from multiple sessions that span a range of parallactic angles. Since our observations were carried out in two sessions that cover different parallactic angles, we examined separately the Stokes $V$ profile from each session, and we found our results to be consistent between sessions. Moreover, \citet{momjian+2012}\ measured the Zeeman effect in the 44 GHz Class I \meth\ maser line toward OMC-2 before and after the new correlator was installed at the VLA, and got the same result for $z\Blos$, despite a change in the maser intensity. Therefore, the observed Stokes $V$ profile is likely not due to instrumental effects. Besides instrumental effects, however, there are other possibilities for non-Zeeman contributions to the Stokes $V$ profile. \citet{wiebe+1998} found that strong linearly polarized maser radiation, together with changes in the orientation of the magnetic field along the line of sight, can produce circular polarization that is, in practice, indistinguishable from that produced by the Zeeman effect. To date, there are no linear polarization measurements of 44 GHz \meth\ masers in the literature. \citet{wiesemeyer+2004} observed the polarization of \meth\ masers at mm wavelengths (84.5 to 157 GHz) and found linear polarization as high as 40\% in one of their sources. Absent measurements of the linear polarization of 44 GHz \meth\ masers, we cannot rule out that linear polarization is responsible for the observed Stokes $V$ profile. However, if linear polarization were responsible, it appears unlikely that it would be converted to circular polarization for the second strongest maser in our field but not the strongest maser. Another issue was described by \citet{houde+2014}, who found that observed antisymmetric and symmetric spectral profiles in SiO masers can arise when the maser radiation scatters off populations of foreground molecules located outside the velocity range covered by the maser. Due to this, the magnetic fields traced by SiO masers are likely much smaller than what would result from a Zeeman interpretation of their observed Stokes $V$ profiles. If future measurements of the Zeeman splitting factor for methanol masers suggest fields that are much larger than plausible, studies of the applicability of the phenomenon discussed by \citet{houde+2014} to Class I \meth\ masers would be important to consider. Finally, another potential issue described in \citet{vlemmings+2011} is that a rotation of the axis of symmetry for the molecular quantum states could cause an intensity-dependent circular polarization that could be mistakenly attributed to a Zeeman splitting. This happens when the maser stimulated emission rate $R$ becomes larger than the Zeeman frequency shift $g\Omega$. In our observation of \droh, $g\Omega \approx 50$~s$^{-1}$. The stimulated emission rate, as written by \citet{vlemmings+2011} is
\begin{equation}
R \simeq \frac{AkT_b \Delta\Omega}{4 \pi h \nu} \protect\label{eq2}
\end{equation}
where $A = 0.392 \times 10^{-6}$~s$^{-1}$ is the Einstein coefficient for the 44 GHz \meth\ maser transition \citep{cragg+93}, $T_b$ is the maser brightness temperature, $\Delta\Omega$ is the maser beaming solid angle, $\nu =$ 44069.488 MHz is the frequency of the maser transition, $k$ is the Boltzmann constant, and $h$ is the Planck constant. For the 44 GHz \meth\ maser in which we have detected the Zeeman effect in \droh, we get $T_b = 4.2 \times 10^6$~K, and from \citet{slysh+2009}, we calculate $\Delta\Omega = 2.0 \times 10^{-2}$. Putting all this information in equation~(\ref{eq2}), we obtain $R \approx 10^{-3}$~s$^{-1}$. Therefore, it is clear that $R \ll g\Omega$, implying that such a non-Zeeman contribution to the observed splitting is unlikely. Moreover, this effect would cause an intensity-dependent polarization, but our strongest maser does not show a Zeeman effect. 

From observations of the Zeeman effect in the CN(1$-$0) line, \citet{falgarone+2008} measured \Blos\ = 0.36 and 0.71~mG in the 1.4 mm sources MM1 and MM2 \citep{woody+89} in \droh, in which the molecular hydrogen densities are $n \sim 1.7 \times 10^5$ cm$^{-3}$. We can use their magnetic field strengths and density, together with the predicted densities for 44 GHz Class I \meth\ masers, to estimate a value for \Blos\ in the 44 GHz line if  the $B \propto n^{1/2}$ relation from \citet{crutcher+99} applies for Class I \meth\ masers. Recently, \citet{leurini+2016} found that bright Class I methanol masers are pumped in high-density regions, corresponding to $n \sim 10^{7-8}$~cm$^{-3}$. Using all of the above information in the $B \propto n^{1/2}$ relation, we find that magnetic field strengths traced by 44 GHz Class I masers should be in the range of 9-17~mG if the 44 GHz \meth\ masers are excited at $n \sim 10^8$~cm$^{-3}$, or as low as 3-5~mG for $n \sim 10^7$~cm$^{-3}$. Meanwhile, \citet{houde+2016} applied a dispersion analysis to polarimetry data observed with the Combined Array for Research in Millimeter Astronomy (CARMA) by \citet{hull+2014} and found the plane of sky magnetic field strength in \droh\ to be $\sim$1.2~mG. If $n \sim 2 \times 10^6$~cm$^{-3}$ in these regions  \citep{houde+2016}, then the $B \propto n^{1/2}$ relation would imply that the magnetic field strengths traced by 44 GHz Class I \meth\ masers are in the range of 3-9~mG if such masers are excited at $n \sim 10^{7-8}$~cm$^{-3}$. While it is possible that the $B \propto n^{1/2}$ relation might not hold in shocked regions, line of sight magnetic field strengths $\sim 10$~mG are certainly plausible in such regions. Using our measured value of $z\Blos = 53.5$~Hz would then imply a Zeeman splitting factor of $z \sim 5$ Hz~mG$^{-1}$ for the 44 GHz methanol line. Such small values of $z \sim 5$ Hz~mG$^{-1}$ would not be surprising because the \meth\ molecule is non paramagnetic, like H$_2$O. For comparison, $z = 2.1$~Hz~mG$^{-1}$ for the 22 GHz H$_2$O maser line \citep{nedoluha+92, sarma+2001}; such values for H$_2$O and \meth\ are a factor $10^3$ smaller than $z = 2.8$~Hz~$\mu$G (i.e., 2800 Hz~mG$^{-1}$) for the 21 cm atomic hydrogen line. Of course, this is only an empirical attempt to determine $z$, and we need more data points in light of the fact that there are no lab measurements or theoretically calculated values of $z$ for the 44 GHz \meth\ transition.

If \Blos\ $\sim$10~mG in these shocked regions, then the magnetic energy density given by $B^2/8\pi$ is $\sim 1 \times 10^{-5}$~ergs cm$^{-3}$, where we have used $B^2 = 3 \Blos^2$ from \citet{crutcher+99}. Meanwhile, the (thermal plus turbulent) kinetic energy density is given by $(3/2)\, mn \sigma^2$, where $m = 2.8\, m_p$ assuming 10\% He, and $m_p$ is the proton mass. The velocity dispersion $\sigma$ is related to the linewidth $\Delta v$ by $\sigma = \Delta v/(8 \ln 2)^{1/2}$. To judge the relevance of the magnetic field, it is best to use a value of $\Delta v$ that is larger than the 0.5~\kmS\ we get from the \meth\ masers, e.g., $\sim$5 \kmS\ measured in the millimeter lines of methanol (at 218 GHz) by \citet{zapata+2012}. With $\Delta v \sim$5~\kmS\ in regions with $n = 10^{8}$~cm$^{-3}$, the kinetic energy density is $3 \times 10^{-5}$~ergs~cm$^{-3}$. That would imply that the magnetic energy density is comparable to the kinetic energy density. Therefore, the magnetic field would have a significant role in the dynamics of these shocked regions even if their densities were as high as $10^{8}$~cm$^{-3}$. Of course, we have to be careful not to get into a circular argument. The comparison of magnetic and kinetic energy densities above has only been made to check for plausibility of the derived field value if the value of $z$ were found to be $\sim 5$ Hz~mG$^{-1}$, as we have determined by the empirical method described above.

\section{Conclusion} \label{conc}

We have detected the Zeeman effect in the 44 GHz Class I \meth\ maser line toward the star forming region \droh. Toward the second brightest maser in \droh,  we find $z\Blos = 53.5 \pm 2.7$~Hz. We have left our result in terms of $z\Blos$ because the Zeeman splitting factor $z$ has never been measured for the 44 GHz transition of \meth. The distribution of the 24 masers that we detected reveals the double arc structure in the redshifted western lobe reported by \citet{araya+2009}. The maser in which we have detected the Zeeman effect is located in the outer arc of this redshifted western lobe, whereas the strongest maser in our observations is located in the inner arc. We did not detect the Zeeman effect in the strongest maser in \droh\ which is located 11\arcsec\ from the second strongest maser in which we detected the Zeeman effect. This is not surprising, given the complex morphology of the plane-of-sky magnetic field as revealed by polarimetric observations with the SMA, in which the plane-of-sky magnetic field vectors were observed to rotate by 90\arcdeg\ over 10\arcsec\ scales \citep{girart+2013}. Zeeman effect measurements in the CN (1$-$0) line \citep{falgarone+2008}, together with the $B \propto n^{1/2}$ relation from \citet{crutcher+99}, imply that magnetic field strengths in the 44 GHz Class I \meth\ maser regions should be $\sim$10~mG; this is also the case if we start with plane-of-sky magnetic field strengths determined using a dispersion analysis of polarimetry data \citep{houde+2016}. Our detected value of $z\Blos$ then implies that $z \sim 5$ Hz~mG$^{-1}$ for the 44 GHz methanol transition; such a small value of $z$ is entirely plausible since the \meth\ molecule is non paramagnetic, like H$_2$O. The observations reported in this paper, therefore, allow for an empirical determination of $z$. Such efforts are important because there are currently no laboratory measurements or theoretically calculated values of $z$ for the 44 GHz \meth\ transition, but data from observations of a larger number of sources are needed to make such empirical determinations robust.

\acknowledgments

We are very grateful to Eric Greisen of NRAO for writing the ZEMAN task in AIPS, and the enhancements to the Gaussian fitting task XGAUS, together with the plotting capabilities introduced through the new task XG2PL. We would also like to express our gratitude to Bill Cotton and Vivek Dhawan of NRAO for helpful discussions and assistance with the AIPS task FRMAP. Finally, we would like to offer our sincere thanks to an anonymous referee for a careful reading and insightful comments on the manuscript that have led to significant improvements.

\facility{VLA}

\clearpage

\begin{deluxetable}{lcrrrrrrrrcrl}
\tablenum{1}
\tablewidth{0pt}
\tablecaption{PARAMETERS FOR VLA OBSERVATIONS
	\protect\label{tOP}}
\tablehead{
\colhead{Parameter} & 
\colhead{Value} }
\startdata
Date \dotfill & 2012 Apr 25    \\
Configuration \dotfill & C \\
R.A.~of field center (J2000) \ldots & 20$^{\text{h}}$~39$^{\text{m}}$~00.8$^{\text{s}}$ \\
Dec.~of field center (J2000) \dotfill & 42\arcdeg~22\arcmin~47\rlap{\arcsec}.\,0\\
Total bandwidth (MHz) \dotfill & 1.0 \\
No.~of channels \dotfill & 256 \\
Channel spacing (km~s$^{-1}$) \dotfill & 0.0266 \\
Approx.~time on source \dotfill & 2 $\times$ 80~min \\
Rest frequency (GHz) \dotfill  & 44.069488 \\
FWHM of synthesized beam \dotfill & $ 0\, \rlap{\arcsec}.\, 63 \times 0\, \rlap{\arcsec}.\, 50$ \\
& P.A. = 75.3\arcdeg \\  
Line rms noise (mJy~beam$^{-1}$) \tablenotemark{a} & 12 \\
\enddata
%\tablecomments{}
\tablenotetext{a}{The line rms noise was measured from the Stokes $I$ image cube using maser line free channels.}
\end{deluxetable}

\clearpage

\begin{deluxetable}{cccccrrrrrcrl}
\tablenum{2}
\tablewidth{0pt}
\tablecaption{FITTED AND DERIVED PARAMETERS FOR THE ZEEMAN DETECTION IN DR21(OH)
	\protect\label{tGauss}}
\tablehead{
\colhead{} &
\colhead{Intensity\tablenotemark{a}} &
\colhead{Center Velocity} &
\colhead{Velocity Linewidth\tablenotemark{b}} & 
\colhead{z\Blos\tablenotemark{c}} \\
\colhead{} &
\colhead{(Jy~beam$^{-1}$)}  &
\colhead{(km~s$^{-1}$)} &
\colhead{(km~s$^{-1}$)} &  
\colhead{(Hz)}}
\startdata
Component 1 & $219.23\pm3.91$  & $0.826\pm0.002$ & $0.365\pm0.002$ & $53.5\pm2.7$     \\
Component 2 & \phantom{1}$82.74\pm2.13$ & $0.531\pm0.009$ & $0.484\pm0.010$ & $-9.2\pm9.3$
\enddata
%\tablecomments{}
\tablenotetext{a}{The intensity values are primary beam corrected.}
\tablenotetext{b}{The velocity linewidth was measured at full width at half maximum (FWHM).}
\tablenotetext{c}{z is the Zeeman splitting factor and \Blos\ is the line of sight magnetic field strength.}
\end{deluxetable}

\clearpage

\begin{deluxetable}{ccccc}
\tablenum{3}
\tablewidth{0pt}
\tablecaption{FITTED AND DERIVED PARAMETERS OF THE OBSERVED MASERS IN DR21(OH)
	\protect\label{tGauss_allmasers}}
\tablehead{
\colhead{R.A. (J2000)} &
\colhead{Dec. (J2000)} &
\colhead{Intensity\tablenotemark{a}} &
\colhead{Center Velocity} &
\colhead{Velocity Linewidth\tablenotemark{b}} \\
\colhead{} &
\colhead{} &
\colhead{(Jy~beam$^{-1}$)}  &
\colhead{(km~s$^{-1}$)} &
\colhead{(km~s$^{-1}$)}}
\startdata
20 38 59.25 & 42 22 48.2  & $ 219.23 \pm 3.91$ & $ 0.826 \pm 0.002$ &   $0.365 \pm 0.002$ \\
     ---    &   ---       & $  82.74 \pm 2.13$ & $ 0.531 \pm 0.009$ &   $0.484 \pm 0.010$ \\  
20 38 59.29 & 42 22 47.0  & $   7.11 \pm 0.12$ & $-0.855 \pm 0.003$ &   $0.372 \pm 0.007$ \\
20 38 59.31 & 42 22 49.1  & $  16.33 \pm 0.12$ & $ 1.099 \pm 0.001$ &   $0.352 \pm 0.003$ \\   
20 38 59.33 & 42 22 46.8  & $   6.64 \pm 0.10$ & $-1.004 \pm 0.004$ &   $0.300 \pm 0.008$ \\
20 38 59.70 & 42 22 41.7  & $   0.35 \pm 0.01$ & $ 0.010 \pm 0.010$ &   $0.786 \pm 0.027$ \\
20 38 59.71 & 42 22 45.3  & $   0.52 \pm 0.04$ & $-0.101 \pm 0.014$ &   $0.402 \pm 0.034$ \\
20 38 59.71 & 42 22 49.1  & $   1.03 \pm 0.34$ & $ 0.202 \pm 0.010$ &   $0.264 \pm 0.041$ \\
     ---    &    ---      & $   0.49 \pm 0.15$ & $-0.036 \pm 0.128$ &   $0.475 \pm 0.148$ \\
20 38 59.76 & 42 22 44.7  & $   0.72 \pm 0.04$ & $-0.249 \pm 0.010$ &   $0.365 \pm 0.025$ \\
20 38 59.85 & 42 22 45.4  & $  14.79 \pm 0.09$ & $-0.216 \pm 0.001$ &   $0.219 \pm 0.003$ \\
20 38 59.85 & 42 22 45.7  & $   3.80 \pm 0.04$ & $-0.510 \pm 0.002$ &   $0.227 \pm 0.004$ \\
20 38 59.89 & 42 22 44.9  & $  16.46 \pm 0.12$ & $ 0.312 \pm 0.002$ &   $0.241 \pm 0.004$ \\
20 38 59.91 & 42 22 44.5  & $  18.11 \pm 0.14$ & $ 0.561 \pm 0.001$ &   $0.203 \pm 0.001$ \\
20 38 59.96 & 42 22 34.7  & $   0.75 \pm 0.05$ & $-1.836 \pm 0.009$ &   $0.267 \pm 0.021$ \\
20 39 00.15 & 42 22 47.3  & $  10.48 \pm 0.07$ & $-1.452 \pm 0.001$ &   $0.387 \pm 0.003$ \\
20 39 00.23 & 42 22 45.4  & $ 323.20 \pm 1.22$ & $ 0.396 \pm 0.001$ &   $0.309 \pm 0.002$ \\       
     ---    &    ---      & $  44.30 \pm 1.22$ & $ 0.098 \pm 0.008$ &   $0.309 \pm 0.011$ \\
20 39 00.25 & 42 22 46.8  & $  11.79 \pm 0.10$ & $ 0.095 \pm 0.001$ &   $0.279 \pm 0.003$ \\
20 39 00.25 & 42 22 46.0  & $   0.83 \pm 0.05$ & $ 1.240 \pm 0.016$ &   $0.605 \pm 0.041$ \\
     ---    &    ---      & $   7.99 \pm 0.06$ & $ 0.367 \pm 0.001$ &   $0.368 \pm 0.003$ \\
20 39 00.30 & 42 22 46.6  & $   0.90 \pm 0.01$ & $-2.049 \pm 0.003$ &   $0.312 \pm 0.007$ \\
20 39 00.33 & 42 22 47.8  & $   1.15 \pm 0.01$ & $-0.029 \pm 0.002$ &   $0.455 \pm 0.004$ \\
20 39 00.51 & 42 22 47.1  & $   0.72 \pm 0.03$ & $ 1.352 \pm 0.009$ &   $0.270 \pm 0.018$ \\
20 39 00.53 & 42 22 47.1  & $   1.82 \pm 0.04$ & $ 2.524 \pm 0.012$ &   $0.468 \pm 0.023$ \\
     ---    &   ---       & $   1.09 \pm 0.19$ & $ 2.116 \pm 0.025$ &   $0.300 \pm 0.054$ \\
20 39 00.52 & 42 22 47.1  & $   2.77 \pm 0.05$ & $ 1.783 \pm 0.009$ &   $0.367 \pm 0.022$ \\
20 39 01.01 & 42 22 41.1  & $   0.81 \pm 0.01$ & $-1.017 \pm 0.002$ &   $0.470 \pm 0.005$ \\
20 39 01.20 & 42 22 40.5  & $   0.63 \pm 0.02$ & $-2.369 \pm 0.011$ &   $0.755 \pm 0.031$ \\
\enddata
%\tablecomments{}
\tablenotetext{a}{The intensity values are primary beam corrected.}
\tablenotetext{b}{The velocity linewidth was measured at full width at half maximum (FWHM).}
\end{deluxetable}

\clearpage
%FIGURES

% Figure 1

% Figure 2

% Figure 3

\end{document}